\documentstyle{grf1996}
\date{March 1996}

\begin{document} 

\bigskip

\begin{frontmatter}

\title{Gravitationally Induced Neutrino--Oscillation Phases}

\author{D. V. Ahluwalia}
\footnote{E-mail address: ahluwalia@nus.lampf.lanl.gov .}
\address{Mail Stop H--846, Physics (P--25) and Theory (T--5) Division \\ 
Los Alamos National Laboratory, Los Alamos, NM 87545, USA}

\author{C. Burgard}
\footnote{E-mail address: christoph.burgard@cern.ch .}
\address{
Universit\"at Hamburg/DESY,
II. Institut f\"ur Experimentalphysik\\
Notkestr.85,
D-22607 Hamburg, Germany}

\begin{abstract}

In this essay, we introduce a new effect of
gravitationally induced quantum mechanical phases
in neutrino oscillations. These
phases arise from an hitherto
unexplored interplay of gravitation and the principle
of the linear superposition of quantum mechanics.
In the neighborhood
of a $1.4$ solar--mass neutron star,
gravitationally induced quantum mechanical phases  
are roughly $20 \%$ of their kinematical counterparts. When this information
is coupled with the mass square differences
implied by the existing 
neutrino--oscillation data we find that the new effect 
may have profound consequences for  type-II supernova evolution.

\end{abstract}

\end{frontmatter}
%\newpage

Neutrinos \cite{WP,RC} play a profound role on almost all length scales.
These scales range from the evolution of the nuclei to the 
evolution of biological structures \cite{JIC}, to the evolution of
the stars, the galaxies, and  the universe \cite{SW,KP}.
The detection \cite{1987a} of an ancient  neutrino 
burst on February 23, 1987
provided a dramatic confirmation of the essential elements of
the physics of type-II supernovae \cite{SAG} and placed
an upper limit of about $10\,\,\mbox{eV}$ on the mass of the electron neutrino
\cite{MR}.
The detection of solar neutrinos in terrestrial detectors provides 
a similarly convincing confirmation of our understanding of the 
solar evolution \cite{JNB}. These important observational
triumphs mark the beginning of a new era: The era of neutrino 
astronomy. The observations on the  solar neutrinos have
already given birth to
the
more controlled accelerator-- and reactor--based neutrino oscillation 
experiments.

For decades the solar
neutrino anomaly \cite{Solar}
has indicated that the neutrino 
flavor eigenstates 
may be a linear superposition of
mass eigenstates \cite{BP}. This view has been further strengthened by the 
data on  atmospheric neutrinos \cite{Kamioka}, and most recently by the 
observation of excess $\overline \nu_e$ 
observed at the Liquid Scintillator Neutrino
Detector (LSND)
Neutrino Oscillation Experiment (NOE)
 at LAMPF \cite{LSND}. The excess events 
observed at LSND NOE
have been tentatively interpreted as  
$\overline{\nu}_\mu\rightarrow
\overline{\nu}_{{e}}$ oscillations \cite{LSND}.

At the present time, (a) the question of neutrino masses \cite{SPR}, (b)
the  relation of the neutrinos
to the spacetime symmetries as manifested in Dirac 
\cite{PAMDeq}, or Majorana \cite{EM,GR,EMa}, or other fundamental
constructs appropriate to the  neutral particles \cite{DVAnp}, (c)
the question of CP violation in the leptonic sector \cite{CP}, and (d) 
other related issues that extend the 
physics beyond the standard model \cite{MP},
are being probed with intense theoretical and experimental vigor.

Gravitation plays no direct role
on 
neutrino oscillations in the existing literature.\footnote{However, 
see (a) Ref. \cite{MS}, 
where Violation of Equivalence Principle
(VEP) in the context of neutrino oscillation experiments is considered;
(b) the work of Goldman {\em et al.}
 on ``Kaons, Quantum Mechanics, and Gravity'' in Ref.
\cite{GNS} (Ref. \cite{GNS}  also
contains references to many classic works on the subject); and (c)
the work of J. Anandan on the gravitational phase operator \cite{JA}.} 
In this essay
we shall introduce the notion of 
gravitationally induced 
quantum mechanical neutrino--oscillation phases 
and discuss the possibility of their observation 
for type-II supernova.

A bit of history is necessary at this point.
In one of the classic experiments of physics,
 Colella, Overhauser, and Werner
(COW) 
established that quantum mechanics and gravitation, despite the 
well--known conceptual problems, behave in a manner expected for any other
interaction \cite{COW}. Given the fact that the experiment involved thermal
neutrons (non-relativistic quantum realm) and the Earth's gravitational field
(weak gravity), this  may not be too 
unexpected. 
Nevertheless, theoretical
investigation of this elegant experiment provides a deep understanding of
gravitation in the context of quantum mechanics \cite{JJS}.
From a formal point of view the COW experiment studies the effect
of gravitation on
the quantum mechanical evolution of a single--mass eigenstate.
For neutrinos,  when we extend the COW--like considerations to
 the linear superposition of mass eigenstates,
a new gravitationally induced quantum mechanical
effect emerges. 
Despite the similarities between the COW effect and the new effect 
introduced here, there are important conceptual differences between
the two effects.
These differences shall also be enumerated at the  appropriate place
in this essay. In a certain sense (to become precise below), 
the two effects shall be seen to be
complementary.

Let us assume that in the ``creation region,''
${\cal R}_c$, located at ${\vec r}_c$, \footnote{The creation region
${\cal R}_c$ is assumed fixed in the global coordinate system.}
 a weak eigenstate with energy $E_\nu$, denoted by
$\vert\nu_\ell,\,{\cal R}_c\rangle$, is produced 
with the clock set to $t=t_c$. Each of the three neutrino mass eigenstates
shall be represented by $\vert \nu_\imath \rangle$; $\imath=1,2,3$. 
So we have the linear superposition:
\begin{equation}
\vert \nu_\ell, \,{\cal R}_c\rangle =
\sum_{\imath=1,2,3} U_{\ell\imath}\,\vert\nu_\imath\rangle\quad,
\label{nu0}
\end{equation} 
where $\ell=e,\mu,\tau$ represents the  weak flavor 
 eigenstates (corresponding
to electron, muon, and tau 
neutrinos respectively).\footnote{We shall assume that CP
is not violated for the purposes of this analysis.
Neutrinos shall be assumed to be of the Dirac type (for a 
recent analysis of various quantum field theoretic possibilities for the 
description of neutral particles of spin--$1/2$ and higher, and their
relation with space--time symmetries, see Ref. \cite{DVAnp} and references
therein). In addition,
we shall assume  that
both $\nu_\ell$ and $\nu_m$ are relativistic in the frame of
the experimenter.}
The $\vert \nu_1\rangle$, $\vert \nu_2\rangle$, and 
$\vert \nu_3\rangle$ correspond to the three mass eigenstates
of masses $m_1,\, m_2,\,\,\mbox{and}\,\, m_3$, respectively. Under the 
already--indicated assumptions the 
$3\times 3$
unitary mixing matrix $U_{\ell\imath}$
may be parameterized by three angles 
and reads \cite[Eq. 6.21 with the CP phase 
$\delta=0$]{KPbook}:
\begin{equation}
U(\theta,\,\beta,\,\psi)\,=\,
\left(
\begin{array}{ccccc}
c_\theta\,c_\beta &{\,\,}& s_\theta\,c_\beta &{\,\,}& s_\beta \\
-\,c_\theta\,s_\beta\,s_\psi\,-\,s_\theta\,c_\psi
&{\,\,}& c_\theta\, c_\psi\,-\,s_\theta\,s_\beta\,s_\psi
&{\,\,}& c_\beta\,s_\psi\\
-\,c_\theta\,s_\beta\,c_\psi\,+\,s_\theta\,s_\psi
&{\,\,}& -\,s_\theta\,s_\beta\,c_\psi\,-\,c_\theta\,s_\psi
&{\,\,}& c_\beta\,c_\psi
\end{array}\right)\quad,
\label{u}
\end{equation}
where $c_\xi\,=\,\cos(\xi)$, 
$s_\xi\,=\,\sin(\xi)$, with $\xi=\theta, \,\beta,\,\psi$.

At a later time $t=t_d>t_c$ ,
we wish to study the weak flavor eigenstate in the ``detector region,''
${\cal R}_d$, located at ${\vec r}_d$.
\footnote{Like the creation region, the detection region 
${\cal R}_d$ too is fixed in the global coordinate system.}
The neutrino evolution from
${\cal R}_c$  to ${\cal R}_d$ is given by the
expression:
\begin{equation}
\vert\nu_\ell,\,{\cal R}_d\rangle =
\exp\left(-\,{i\over {\hbar }} \int_{{t}_c}^{{t}_d}
H \mbox{dt} \,+\, {i\over \hbar}
\int_{{\vec r}_c}^{{\vec r}_d}
\vec P\cdot d\vec x
\right)
\vert\nu_\ell,\,{\cal R}_c\rangle\quad.
\end{equation}

Here $H$ is the time translation operator
(the Hamiltonian)
 associated with  the system; $\vec P$ is the operator
for spatial translations (the momentum operator), and 
$\left[H(t,\vec x),\,{\vec P}(t,\vec x)\right]=0$.
Consider
 that both ${\cal R}_c$ and ${\cal R}_d$ are located in the 
Schwarzschild
gravitational environment \cite{SW}
of a spherically symmetric object of mass $M$.
The direction  of neutrino propagation 
is along $\vec L = \vec r_d -\vec r_c$.

We now confine to the weak field limit, and 
neglect the spin--dependent terms
for the present, as  we do not wish to study effects
of the astrophysical magnetic fields, or the effects of interaction between
spin (of neutrino) and angular momentum (of the astrophysical object).
Under these conditions, the work of Stodolsky implies \cite{LS}: 
\begin{eqnarray}
\exp && \left(-\,{i\over {\hbar }} \int_{{t}_c}^{{t}_d}
H \mbox{dt} \,+\, {i\over \hbar}
\int_{{\vec r}_c}^{{\vec r}_d}
\vec P\cdot d\vec x
\right)
\vert\nu_\imath\rangle \nonumber\\
&&\qquad\qquad\qquad\qquad\,=\,
\exp\left[-\,{i\over {\hbar }} 
\int_{{\cal R}_c}^{{\cal R}_d}
\left(
\eta_{\mu\nu}+{1\over 2} h_{\mu\nu}\right)
p_\imath^{\mu}\,\mbox{dx}^\nu\right]
\vert\nu_\imath\rangle
\quad,
\end{eqnarray}
where $h_{\mu\nu}=g^W_{\mu\nu}-\eta_{\mu\nu}$;
$g^W_{\mu\nu}$ is the 
Schwarzschild space--time metric in the weak field limit and $\eta_{\mu\nu}$
is the flat space--time metric. 
In addition, $h_{\mu\nu}=2\,\phi\,\delta_{\mu\nu}$ with the dimensionless
gravitational potential $\phi=-\, GM/(c^2\,r)$.
Further, it shall
be 
noted that $p_\imath^{\mu}$ is the four--momentum of special relativity
with $p^\mu\equiv  m\, dx^\mu/ds_0$ in the notation of Ref. \cite{LS}.
The limits of applicability of this formalism are further enumerated in 
the above--cited paper of Stodolsky.
To avoid notational confusion we remind the reader that $p_\imath
\equiv \vert \vec p_\imath\vert$ --- the subscript
$\imath$ identifies the mass eigenstate, and does not refer to $\imath$th
component of the momentum vector.

We now calculate
the ``neutrino oscillation probability'' from a state
$\vert\nu_\ell,\,{\cal R}_c\rangle$ 
to another state $\vert \nu_{\ell^\prime}, \,{\cal R}_d\rangle$
following closely the standard arguments,
appropriately adapted to the present situation \cite{KPbook,HL,TG}.
The oscillation probability
is obtained by calculating the projection
$\langle\nu_{\ell'},\,{\cal R}_d\vert\nu_\ell,\,{\cal R}_c\rangle$,
i.e., the amplitude for 
$\vert \nu_\ell,\,{\cal R}_c\rangle \rightarrow
\,\vert\nu_{\ell'},\,{\cal R}_d\rangle$, and then multiplying it by its
complex conjugate. An 
algebraic exercise that  (a) exploits
the unitarity of the neutrino mixing matrix
$U(\theta,\,\beta,\,\psi)$, (b) exploits orthonormality 
of the mass eigenstates, (c) exploits certain trigonometric identities, 
 and  (d) takes care of the fact
 that
now $dx$ and $dt$ are related by
\begin{equation}
dx\simeq\left[1-\left({{2GM}\over {c^2 r}}\right)\right]\,c \,dt
\quad,\label{xt}
\end{equation}
yields:
%\vbox{
\begin{eqnarray}
{\cal P} \Big[\vert\nu_{\ell},{\cal R}_c\rangle\rightarrow 
\vert\nu_{\ell^\prime},{\cal R}_d\rangle\Big] \,= \,
\delta_{\ell\,\ell'}
 &&\,- \,4\,U_{\ell'\,1}\,U_{\ell\,1}\,U_{\ell'\,2}\,
U_{\ell\,2}\,\sin^2\left[  
\varphi^0_{21} + \varphi^G_{21}
\right]\nonumber\\
&&\,-\,4\,U_{\ell'\,1}\,U_{\ell\,1}\,U_{\ell'\,3}\,
U_{\ell\,3}\,\sin^2\left[
\varphi^0_{31} + \varphi^G_{31}
\right] \nonumber\\
&&\,-\,4\,U_{\ell'\,2}\,U_{\ell\,2}\,U_{\ell'\,3}\,
U_{\ell\,3}\,\sin^2\left[
\varphi^0_{32} + \varphi^G_{32}
\right]\quad.
\label{prob}
\end{eqnarray}
%}
The arguments of $\sin^2(\cdots)$ in the neutrino
oscillation probability
 now contain two types of phases.
The usual {\em kinematic phase}, denoted here by $\varphi^0_{\jmath\imath}$,
and defined as
\begin{equation}
\varphi^0_{\jmath\imath} \equiv {{ c^3}\over{ 4\hbar}}
{{ \vert {\vec r}_d - {\vec r}_c\vert \Delta 
m^2_{\jmath\imath}}\over {E}}
=
{{ c^3}\over{ 4\hbar}}
{{L\, \Delta 
m^2_{\jmath\imath}}\over {E}}
\quad ; \label{phik}
\end{equation}
and the new
 {\em gravitationally induced quantum mechanical phase},
denoted here by $\varphi^G_{\jmath\imath}$, and defined as
\begin{equation}
\varphi^G_{\jmath\imath} \equiv
{G M c\over{4\hbar}}
\left[
\int_{{\vec r}_c}^{{\vec r}_d}
{\mbox{dL}\over r}
\right]
{{\Delta m^2_{\jmath\imath}}\over E}
\quad. \label{phig}
\end{equation}

It is readily seen that the gravitationally induced 
phases and the 
kinematic phases, for neutrino oscillations, are related via the identity
\begin{equation}
\varphi^G_{\jmath\imath} = -\, \langle\phi\rangle \,\varphi^0_{\jmath\imath}\quad,
\end{equation}
where $\langle\phi\rangle$ is the average dimensionless gravitational
potential over the semi-classical neutrino path
\begin{equation}
\langle\phi\rangle \equiv
-\,{1\over L}\,\int_{{\vec r}_c}^{{\vec r}_d} dL\, {{ G M}\over{c^2 r}}
\quad.
\end{equation}

Two immediate questions seem relevant.
First, what are the conceptual
similarities, and differences,  
between the neutron interferometer
experiment of COW  and the phenomenon of neutrino oscillations
in the presence of gravity? 
Second, what are the astronomical, or experimental chances, 
of detecting the gravitationally 
induced modification to neutrino oscillations?
We shall discuss these two questions in turn, and then proceed to 
explore the astrophysical consequences of our study.

The COW experiment  studies the effect of gravitation on a 
{\em single mass eigenstate\/ }
(i.e., neutron of mass $m_n \simeq 940\,\,\mbox{MeV}$). The observable
physical effect arises because {\em spatially distinct} parts (spatial spread
$\simeq {\mbox{a few cm}}$) of the wave
function pick up different gravitationally induced quantum mechanical phases.
In the case of neutrino oscillations 
the gravitationally induced modification to
the $\vert\nu_\ell\rangle\rightarrow\vert\nu_{\ell^\prime}\rangle$
oscillations arises from the 
difference
in phase that {\em each of the different mass eigenstates} 
picks in a given gravitational environment.
{\em No spatial spread} of the wave function is required.
Thus, the formal difference between the 
effect induced by gravity in the COW experiment, and the one induced
in neutrino oscillations, is that in the former 
 the different strengths of 
the gravitational field at different locations (and 
interacting with the different superimposed amplitudes 
of the wave function 
associated with a neutron mass eigenstate) 
manifest into a physically observable result; whereas in the latter  the 
different gravitational interaction energies--momenta of the 
respective mass eigenstates 
leave their trace in the  gravitationally induced quantum  mechanical
phases. The magnitude of the two effects is also affected by
the non-relativistic nature of the neutrons in the COW experiment,
and the extreme relativistic nature of the neutrinos.

An important question 
in the context of our present discussion is 
related to the observability of the new effect. Let us 
again note that
in the COW  experiment the relevant quantity,
apart from Earth's gravitational field, is the product of
mass of neutron and the (vertical)
spatial spread of the wave function. For neutrino mass eigenstate
of the order of an $\mbox{eV}$ (nine orders of magnitude below
the neutron mass) one may expect that if the flavor eigenstate of a neutrino
is allowed to travel a (vertical or horizontal!)
distance of the order of $10^9\,\,\mbox{cm}$
(actually ten to a hundred  times less than this will do because of
the fact that
the COW experiment saw  a shift in fringes by a rather large number) one may
see gravitationally induced effect on neutrino oscillations
in terrestrial experiments (such as those involving atmospheric neutrinos
\cite{Kamioka}).
However, this is not so. The reason is that
the COW experiment used
thermal neutrons (i.e., non-relativistic particles) and hence the effect
was proportional to the  {\em mass} of the neutron. For  the 
$\mbox{MeV}$--$\mbox{GeV}$ neutrinos
(i.e., relativistic particles), as is the case for the accelerator and 
the atmospheric neutrinos, the effect
is proportional to {\em mass  squared differences}.
As a result, the effect may only be
seen in astrophysical environments.

The above considerations complete the formal aspects of this paper.
We now turn to what is essentially a back--of--the--envelope
exploration of the astrophysical consequences.

In the context of supernova explosions, and the problem of obtaining
successful explosions, we 
follow Colgate {\em et al.} \cite{SAG}
 and  assume that the matter next to 
the neutron star is heated by neutrinos from the cooling neutron star.
Colgate {\em    et al.} note that in some models ``this results in strong,
large scale convective flows in the gravitational field of the neutron 
star that can drive successful, albeit weak, explosions.''
Now we recall that 
 the energy flux in each 
of the electron neutrinos and antineutrinos is about
$L_{\nu_e}\approx L_{\overline{\nu}_e}\approx \mbox{few}\times 10^{52} \,\,
\mbox{ergs}\,\, {\mbox s} ^{-1}$,
with comparable fluxes of  $\nu_\mu$, $\overline{\nu}_\mu$,
$\nu_\tau$, $\overline{\nu}_\tau$ \cite{BW}.
 A particularly relevant fact about these fluxes 
is that  while average energy of $\nu_e$ is about $10\,\,\mbox{MeV}$, 
the average
energy of other neutrinos may be higher by a factor of about 
$2$ to $3$.
Because of the extremely large fluxes, and different energies
(and hence different cross sections) associated with the neutrinos,
even a small variation in the neutrino oscillation probabilities may
profoundly affect the success of the supernova explosion provided at 
least one of the 
$\lambda^{\rm osc}_{\jmath\imath}$ has appropriate 
length scale.\footnote{
In neutrino oscillation literature ``oscillation length'' is defined as  
\[
\lambda^{\rm osc}_{\jmath\imath}\,=\, {{2\,\pi}\over\eta}\,{E\over{\Delta 
m^2_{\jmath\imath}}}\quad,\]
where $
{\Delta 
m^2_{\jmath\imath}}$ are measured in $\mbox{eV}^2$, $E$ in $\mbox{MeV}$,
and $\eta = 1.27$.}

The existing experiments imply  
that the two independent mass square 
differences \cite{DVAt} that define $\varphi^0_{\jmath\imath}$ are as follows:
$\Delta m^2_{21} \simeq   10^{-2}\,\, \mbox{eV}^2$
if the zenith--angle  dependence of the atmospheric neutrino
anomaly is explained by neutrino oscillations (or \cite{CF},
$\Delta m^2_{21} \simeq   10^{-5}\,\, \mbox{eV}^2$ if the energy
dependence of the solar neutrino deficit is explained by invoking the
MSW 
 phenomenon \cite{MSW})
 and
$\Delta m^2_{32} \simeq 0.5\,\, \mbox{eV}^2$
(and opposite  these values if the inverted mass hierarchy is considered
\cite{GMF}). 
Taking the typical energy of a supernova neutrino to be $10\,\,
\mbox{MeV}$, these mass square 
differences 
yield oscillation lengths
$\lambda^{\rm osc}_{21} \simeq 5\,\,{\mbox {km}}$ 
(or, $\lambda^{\rm osc}_{21} \simeq 5\times 10^3\,\,{\mbox {km}}$)
and 
 $\lambda^{\rm osc}_{32} \simeq 100\,\,{\mbox m}$.
These length scales are certainly relevant to the
supernova--evolution processes of  {\em neutrino diffusion\/},
{\em neutrino trapping}, and {\em neutrino heating}.\footnote{
For details on neutrino heating in the context of supernova 
explosion we refer the interested reader to Ref. \cite{SAG}, and on 
supernova--evolution processes of  neutrino diffusion and 
neutrino trapping the reader may find
Ref. \cite{BBAL} very valuable.}
These length scales, 
and hence the associated  supernova processes (and very importantly the
compatibility arguments between terrestrial neutrino oscillations and 
supernova evolution \cite{CF}), will be altered if 
we find 
$\varphi^G_{\jmath\imath}$ equals a few percent of
$\varphi^0_{\jmath\imath}$.

To see if the neutrino oscillation phases can be altered at a level 
of a few percent in the neighborhood of the  neutron star we 
 consider 
the radially outward motion of a neutrino, and set
${\vec r}_d = \alpha\, {\vec r}_c$, $1<\alpha\le\infty$, then we find
\begin{equation}
\left(\varphi^G_{\jmath\imath}\right)_{\Vert} = 
{ G M c\over{4\hbar}}
{{\Delta m^2_{\jmath\imath}}\over E}\ln(\alpha) =
 \left[
{ {G M}\over {c^2 r_c}}\,
{\ln{\alpha}\over {\alpha-1}}\right]\, \varphi^0_{\jmath\imath} 
\quad.
\end{equation}
For motion transverse to the radial direction
(and in the {\em vicinity\/} 
of ${\vec r}_c$), the corresponding expression
is 
\begin{equation}
\left(\varphi^G_{\jmath\imath}\right)_{\perp} = 
{ G M c \over{4\hbar}}
{{\Delta m^2_{\jmath\imath}}\over E}
{{ \vert {\vec r}_d - {\vec r}_c\vert}\over {r_c}}
=
\left[{{G M}\over {c^2 r_c}}\right]\,\varphi^0_{\jmath\imath} 
\quad.
\end{equation}

Taking $M$ to be $1.4$ solar mass, $1.4\,M_\odot$, and
$r_c=  10\,\,{\mbox{km}}$, we have
\begin{equation}
\left(\varphi^G_{\jmath\imath}\right)_{\Vert} = \, 0.21 
\left({{\ln(\alpha)}\over{\alpha-1}}\right)
\,\varphi^0_{\jmath\imath},\quad
\left(\varphi^G_{\jmath\imath}\right)_{\perp} = \, 0.21 
\,\varphi^0_{\jmath\imath}\quad.
\end{equation}

We thus find that
for a $1.4$ solar mass neutron star, with a radius of ten kilometers,
$\varphi^G_{\jmath\imath}$ is about twenty percent of the 
$\varphi^0_{\jmath\imath}$. 
In astrophysical situations matter effects, and the
presence of magnetic fields (if neutrinos have non--zero magnetic moments),
will further alter neutrino oscillations \cite{MSW,CPv,BG}. The
gravitationally induced 
neutrino--oscillation phases arise from an hitherto
unexplored interplay of gravitation and the principle
of the linear superposition of quantum mechanics and cannot be ignored
in many astrophysical environments.

\vskip 1.0cm
\noindent
{\em Acknowledgments }
We wish to gratefully 
acknowledge useful comments of Yuval Grossman, and the comments of
Harry Lipkin forwarded by him to us,  on the 
first draft of this manuscript. 
These comments led to a correction
in the sign of the effect introduced here.
We also thank them for providing us with a rough draft of their notes
where, after reading our work, they reproduced our results in a 
covariant analysis.
 In addition,
one of us (DVA) wishes to record the following acknowledgements.
I wish to extend my {\em zimpoic} thanks to Terry Goldman,
Bill Louis, John McClelland,  Hywel White, and Nu Xu
for our continuing conversations  on neutrino oscillations and 
other matters of physics.  With Nu I also realised  that  
a dark photon night is  a bright neutrino day, 
for this too I thank him warmly.
Warm thanks are extended to Jeanne Bowles for many stylistic suggestions.
Affectionate thanks are due to
Marianne Raish for her good cheer 
and help in taking care of non--academic matters, thus
allowing me to devote  myself to these studies.

{\em
This work was done, in part, under the auspices of the 
U. S. Department of Energy; and supported by facilities and
financial support of the groups P-25 and T-5
of the Los Alamos National Laboratory.}

\newpage
\appendix{\bf Appendix A: Erratum}

The following Erratum was published in \textit{Gen. Rel. Grav.} {\bf 29},
681 (1997):

\bigskip

\hrule
\medskip
The $2\hbar$ in eqs. (7), (8), (11), and (12) should be replaced 
by $4\hbar$.

\medskip

\noindent
In retrospect, this paper shows that neutrino oscillations
provide a flavor-oscillation clock \textit{and} this flavor-oscillation
clock redshifts as required by the theory of general relativity.
\medskip
\hrule

\newpage

\appendix{\bf Appendix B: Neutrino oscillations and supernovae}

The following is an \textit{unedited} text of Section 1
of ``JRO Fellowship Research Proposal by D. V. Ahluwalia,''
which was submitted by Mikkel B. Johnson in July 1996 in
his \textit{Nomination of Dharam Ahluwalia for Oppenheimer 
Fellowship.} The appendix title coincides with the 
title of Section 1 of the proposal.

\bigskip
\hrule
\medskip
Neutrinos were introduced in physics by Pauli to save conservation 
of energy and momentum in the $\beta$-decay: Neutron $\rightarrow$ 
Proton + Electron + Anti-electron Neutrino. All the planets and  galaxies are embedded in a sea of neutrinos with a number density of roughly 100 neutrinos/cm$^3$. Our own Sun shines via thermonuclear processes 
that emit neutrinos in enormous number. Because of their weak interactions, neutrinos, unlike photons, can pass through extremely dense matter very efficiently. This fact makes neutrinos primary agents for energy transport 
in the dense matter associated with supernovae and neutron stars.

  Since their initial experimental observation by Frederick Reines 
and C. L. Cowan, neutrinos are now known to exist in three types. 
These types are called ``electron,'' ``muon,'' and ``tau'' and 
are generically written as $\nu_e$, $\nu_\mu$, and $\nu_\tau$. 
A series of empirical anomalies indicates that  neutrinos may not have a definite mass but, instead, be in a linear superposition of three different mass eigenstates. The mass differences in the underlying mass eigenstates would cause a neutrino of one type to ``oscillate'' to a neutrino of another type as may have been seen recently at the LSND neutrino oscillation experiment at LANL. The phenomenon of neutrino oscillations, if experimentally confirmed, will have profound consequences not only for nuclear and particle physics but also for astrophysics and cosmology.

   I have already noted the neutrinos to be prime drivers of supernova explosions. The phenomenon of neutrino oscillations will alter the evolution of supernova explosion. The basic problem that still stands unsolved is a robust theory of supernova explosions. In the context of supernova explosions, and the problem of obtaining successful explosions, I now follow Colgate 
\textit{et al.} 
[S. A. Colgate, M. Herant, and W. Benz, Phys. Rep. {\bf 227}, 157 (1993)] 
and assume that the matter next to the neutron star is heated by neutrinos from the cooling neutron star. They note that in some models ``this result in strong, large scale convective flows in the gravitational field of the neutron star that can drive successful, albeit weak, explosions.'' I emphasize that all authors find that without ``fine tuning'' the explosions are weak and lack about five percent of the energy needed for an explosion. Qualitatively, this missing energy needed for a robust model of explosion may be provided if the length scales over which neutrino oscillations take place are of the same order of magnitude as the spatial extent of a neutron star and neutrino-sphere, because while

\begin{quote}
        the energy flux in each of the electron neutrinos and antineutrinos is 
        about $L_{\nu_e}\approx   L_{\overline{\nu}_e}\approx$   
few$\times$  $10^{52}$   ergs s$^{-1}$, 
with comparable fluxes of $\nu_\mu$, $\overline{\nu}_\mu$,
$\nu_\tau$, and  $\overline{\nu}_\tau$,
\end{quote}
the 
\begin{quote}
        average energy of $\nu_e$  
is about 10 MeV, the average energy of other 
        neutrinos is \textit{higher} 
by a factor of 2 for $\nu_\mu$    and   $\overline{\nu}_\mu$, 
and by a factor of 3 for    $\nu_\tau$, and  $\overline{\nu}_\tau$.
\end{quote}
Any oscillation between neutrinos of different flavors is, therefore, an indirect energy transport mechanism towards the actively interacting
$\nu_e$    and   $\overline{\nu}_e$.
Qualitatively this contributes in the direction of the robustness of the explosion. My resent work, with C. Burgard, on the solution of terrestrial neutrino anomalies provides precisely the neutrino oscillation 
parameters that yield the oscillation length scales of just the right order of magnitude for supernova physics (and in addition predict the observed solar neutrino deficit).

   In order to make these qualitative arguments quantitative two 
additional physical 
processes affecting the above indicated \textit{vacuum}
neutrino oscillations must be incorporated: (a) The presence of large electron densities in astrophysical environment makes it necessary that relevant matter induced effects, suggested by Mikheyev, Smirnov and 
Wolfenstein, be considered, 
and (b) My work, with C. Burgard, on gravitationally induced neutrino oscillation phases also indicates  that strong gravitational fields associated with neutron stars may introduce important modifications to neutrino oscillations, and hence to the suggested energy transport mechanism via neutrino oscillations. As part of my JRO studies I propose to implement the above outlined program quantitatively. My quantitative and qualitative studies so far give reasons to claim that there is every 
 physical reason to believe that the ``missing energy'' in the non-robust models for supernova explosion, the anomaly in the observed deficit in the solar neutrino flux, the excess 
$\overline{\nu}_e$
  events seen at 
LSND at Los Alamos, and the anomaly associated with atmospheric neutrinos, 
\textit{all} 
arise from the same underlying new physics \textemdash 
the phenomenon of neutrino oscillations from one type to another. It is of profound physical importance to place these suspected physical connections on firm quantitative foundations.

\medskip
\hrule

\bigskip
The above proposal has been widely known informally (without a full
access to its text). This appendix fills the gap 
of its availability.

The above suggestion  has been pursued vigorously, 
though often without being acknowledged. 
Calculations done by S. Goswami,  
Pramana {\bf 54}, 173-184 (2000) [arXiv: hep-ph/0104094] 
strongly support the the suggestion that neutrino 
oscillations are indeed a powerful energy transport mechanism 
and may play a dominant and important role in 
supernova explosions.

\end{document}